\documentclass[10pt, final, conference, twoside]{IEEEtran}
\usepackage{cite}
\usepackage{amsmath,amssymb,amsfonts}
\usepackage{array,multirow,graphicx}
\usepackage{tikz}
\usepackage{scalerel}
\usepackage{mathtools,eqparbox}
\usepackage{xcolor}
\usepackage{commath}
\usepackage{pgfplots}
\usepackage{subfigure}
\usepackage[hidelinks]{hyperref}
\usepackage[utf8]{inputenc}
\usepackage[T1]{fontenc}
\usepackage{psfrag}
\usepackage{flushend}
\def\Var{{\textrm{Var}}}
\def\E{{\textrm{E}}}

\newcommand{\C}{\cal{C}}
\newcommand{\rg}{R_{g}}

\newcommand{\ru}{R_{u}}

\newcommand{\lambdahaps}{\lambda_\mathrm{{HAP}}}
\newcommand{\lambdaris}{\lambda_\mathrm{{RIS}}}

\newcommand{\HRIS}{H_{\mathrm{RIS}}}
\newcommand{\HHAP}{H_{\mathrm{HAP}}}

\def\Var{{\textrm{Var}}}
\def\E{{\mathbb{E}}}

\newtheorem{optimization problem}{Optimization Problem}
\pgfplotsset{compat=1.14}

\tikzstyle{arrow} = [thick,->,>=stealth]
\tikzstyle{block} = [rectangle, rounded corners, minimum width=1cm, minimum height=1cm,text centered, draw=black, fill=red!30]
\tikzstyle{input} = [circle, minimum width=2.5cm, minimum height=1cm, text centered, draw=black, fill=blue!30]

\usetikzlibrary{svg.path}
\definecolor{orcidlogocol}{HTML}{A6CE39}
\tikzset{
    orcidlogo/.pic={
        \fill[orcidlogocol] svg{M256,128c0,70.7-57.3,128-128,128C57.3,256,0,198.7,0,128C0,57.3,57.3,0,128,0C198.7,0,256,57.3,256,128z};
        \fill[white] svg{M86.3,186.2H70.9V79.1h15.4v48.4V186.2z}
        svg{M108.9,79.1h41.6c39.6,0,57,28.3,57,53.6c0,27.5-21.5,53.6-56.8,53.6h-41.8V79.1z M124.3,172.4h24.5c34.9,0,42.9-26.5,42.9-39.7c0-21.5-13.7-39.7-43.7-39.7h-23.7V172.4z}
        svg{M88.7,56.8c0,5.5-4.5,10.1-10.1,10.1c-5.6,0-10.1-4.6-10.1-10.1c0-5.6,4.5-10.1,10.1-10.1C84.2,46.7,88.7,51.3,88.7,56.8z};
    }
}

\newcommand\orcidicon[1]{\href{https://orcid.org/#1}{\mbox{\scalerel*{
                \begin{tikzpicture}[yscale=-1,transform shape]
                \pic{orcidlogo};
                \end{tikzpicture}
            }{|}}}}
\newcounter{MYtempeqncnt}

\IEEEoverridecommandlockouts

\begin{document}

\title{Integrating RIS into HAP Networks\\for Improved Connectivity\\
}

\author{Islam~M.~Tanash\IEEEauthorrefmark{1}${\textsuperscript{\orcidicon{0000-0002-9824-6951}}}$, Ayush Kumar Dwivedi\IEEEauthorrefmark{2}${\textsuperscript{\orcidicon{0000-0003-2395-6526}}}$, and Taneli Riihonen\IEEEauthorrefmark{2}${\textsuperscript{\orcidicon{0000-0001-5416-5263}}}$\\
\IEEEauthorrefmark{1}Department of Electrical Engineering, Prince Mohammad Bin Fahd University, Saudi Arabia\\
\IEEEauthorrefmark{2}Faculty of Information Technology and Communication Sciences, Tampere University, Finland\\
\texttt{itanash@pmu.edu.sa}, \texttt{ayush.dwivedi@tuni.fi}
}
\maketitle

\begin{abstract}
This paper investigates a high-altitude platform (HAP) network enhanced with reconfigurable intelligent surfaces (RISs). The arbitrary placement of HAPs and RISs is modeled using stochastic geometry, specifically as homogeneous Poisson point processes. The HAP--RIS links are assumed to follow Rician fading, while the RIS--user links experience shadowed-Rician fading. The system's coverage probability and ergodic capacity are derived analytically and validated through Monte Carlo simulations. The results highlight significant performance gains and demonstrate the influence of various system parameters and fading conditions. The proposed system has potential for enhancing connectivity and data offloading in practical scenarios.%
\end{abstract}


\section{Introduction}
The demand for wireless services is growing due to the rise of mobile broadband and the Internet-of-Things (IoT). This has led to increased use of non-terrestrial networks (NTN), including satellite and high-altitude platform (HAP) networks \cite{satelliteINGR, Dwivedi_IoTJ}. Recently, HAPs have gained popularity because of their agility \cite{Alam-MCOM2021:High,Alfattani-MVT23:Multimode, ITU-HAPS}. These stratospheric platforms, operating at altitudes between 20--50 km, offer the potential to provide wide-area coverage with lower latency compared to satellite systems and greater flexibility than terrestrial networks \cite{satelliteINGR}. They can rapidly deploy broadband services to underserved areas, support disaster relief operations, and augment existing communication infrastructure. 

However, the effectiveness of HAP-based networks faces challenges in diverse environments. Urban environments often suffer from significant blockages and signal degradation due to buildings, while in rural areas, maintaining reliable connectivity over vast terrain and distances can be difficult. These factors collectively strain HAP systems' capacity and energy resources, challenging their ability to provide reliable, high-quality service.

The reconfigurable intelligent surface (RIS) is an emerging technology capable of overcoming the aforementioned limitations \cite{Tekbıyık-MVT22:Reconfigurable}. They are passive, programmable surfaces that dynamically reflect wireless signals in desired directions, creating a virtual line-of-sight (LoS) path between a HAP and end-users. By strategically integrating RISs into HAP networks, it is possible to mitigate the effects of blockages, reduce signal attenuation, and improve coverage reliability. Various architectures exist to incorporate RISs with HAPs, including placing RISs on aerial platforms or using terrestrial RISs.

This study focuses on a terrestrial RIS (TRIS)-enabled HAPs network, where the RISs are deployed on ground-based structures to overcome coverage challenges \cite{Ye-JPROC22:Nonterrestrial}. Such TRIS can function as versatile infrastructure, supporting HAP-based and terrestrial wireless communications, thus enhancing system reusability. The proposed system model targets a wide range of use cases, including enhanced mobile broadband in urban centers, reliable connectivity for IoT devices in smart cities, and improved coverage for rural and remote areas.

Recent research has explored deployment strategies, operational altitude, and coverage capabilities for HAPs \cite{Abbasi-MWC24:HAPS}. Parallel studies on RISs have demonstrated their capacity to enhance signal strength and reliability in NTNs \cite{Ye-JPROC22:Nonterrestrial,Tekbıyık-MVT22:Reconfigurable}. The integration of RISs with terrestrial networks and NTNs is analyzed in \cite{Ramezani-NET2022:Toward}, particularly in the context of a NOMA-based system. Similarly, a multi-layer RIS-assisted receiver was used to overcome severe fading and energy limitations in the HAPs network, significantly improving system performance through a robust optimization framework \cite{Kang-TWC2024:Exploting}. A resource-efficient HAP--RIS system that optimizes energy consumption and improves QoS through intelligent resource management was proposed in \cite{Alfattani-LWC23:Resource-Efficient,Alfattani-GCWkshps22:Beyond-Cell}, enhancing beyond-cell communications. A multi-objective optimization approach for aerodynamic HAPs equipped with a RIS was presented in \cite{Azizi-LWC23:RIS}, addressing delay, trajectory, and channel variations to optimize performance in dynamic environments. Additionally, the performance of a RIS-assisted HAPS relaying system was presented in \cite{HAP_RIS_ICCSPA} using stochastic geometry to model the location of HAPs, RISs, and obstructing buildings. On the standardization front, solutions for 5G New Radio (NR) supporting airborne platforms were presented in the 3rd Generation Partnership Project (3GPP) \cite{3GPP-TR38.821:Solutions_NTN}, which became normative requirements for NTNs in Release 17. Meanwhile, ETSI's industry specification group has studied RISs' use cases, deployment scenarios, and standardization impact \cite{ISG_RIS_ETSI}. 

We propose a system concept for a terrestrial RIS-enabled HAP network, connecting users to the HAP through the nearest RIS. Tools from stochastic geometry are applied to randomly model the locations of HAPs and RISs by homogeneous Poisson point process (PPP), thus capturing the essential characteristics of the system in a generalized manner. Unlike our previous work in \cite{HAP_RIS_ICCSPA}, which used a specific Boolean model for an urban environment, this study models a generic environment using shadowing. 

We derive closed-form expressions for coverage probability and ergodic capacity, assuming Rician fading between a HAP and a RIS and shadowed-Rician fading between a RIS and a user. These analytical results provide a robust framework for evaluating system performance under realistic channel conditions. Finally, we generate insights into the impact of channel fading and critical system parameters, such as the height of the HAP network and intensity of RIS deployment, offering valuable guidance for designing and optimizing future RIS-enabled HAP networks.
\begin{figure}[!t]
    \centering
    \includegraphics[width=\linewidth]{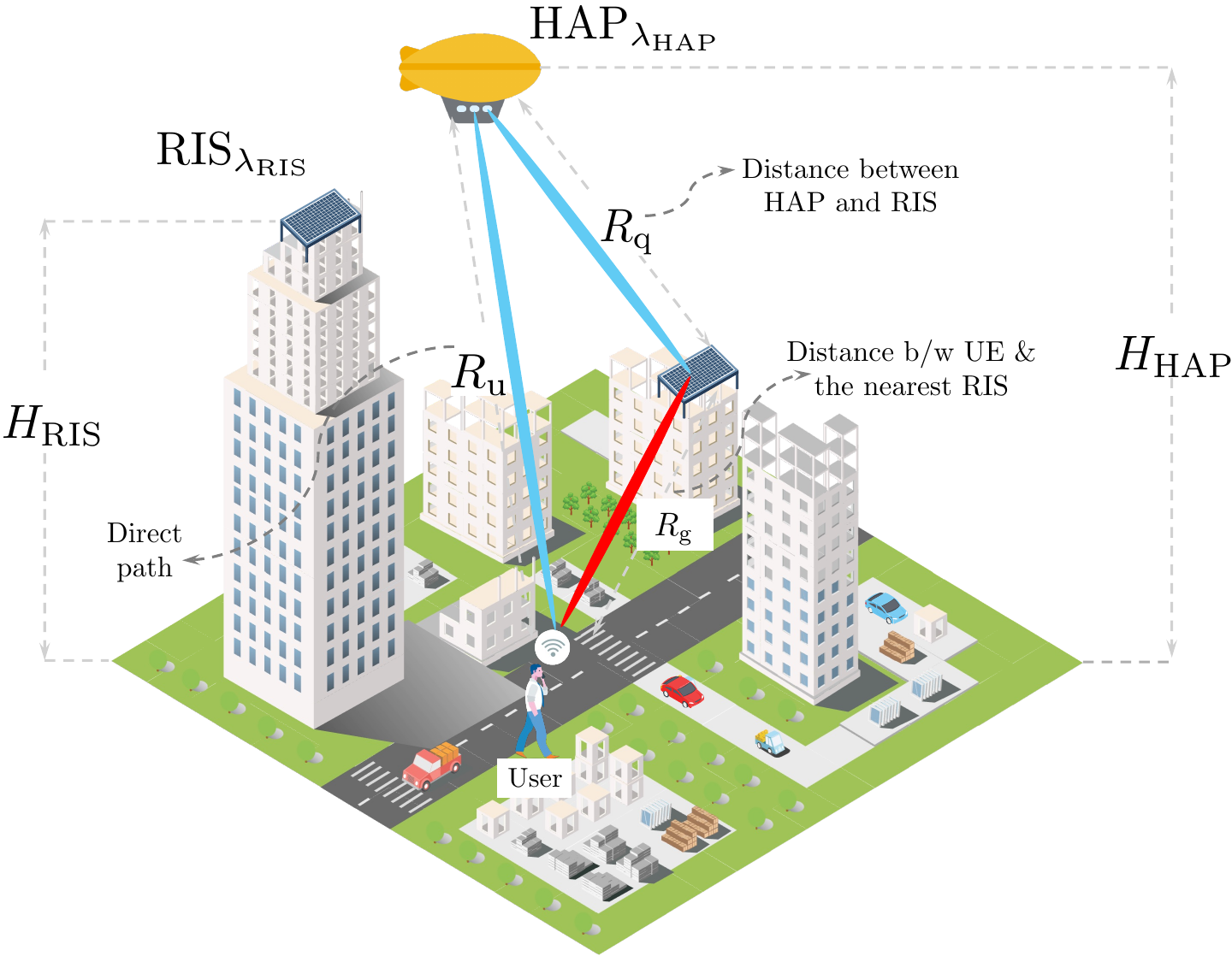}
    \caption{The proposed system comprises a network of HAPs integrated with RISs to establish virtual LoS communication. The user receives a reflected signal from RIS and a weak direct signal from the HAP. The locations of both the HAPs and the RISs are modeled randomly according to a PPP.}
    \label{fig:systemModel}
\end{figure}

\section{System Model}
We consider a network of HAPs deployed in a two-dimensional (2D) space at an altitude $\HHAP$. The deployment follows a homogeneous PPP on the $\mathbb{R}^2$ plane with a density of $\lambdahaps$. 
Due to the high altitude of the HAPs, ranging from $20$ km to $50$ km, the signal received by users can be significantly attenuated. To mitigate this, RISs are introduced to enhance system performance by improving signal propagation and coverage. The RISs are assumed to be deployed at a height $\HRIS$, following a homogeneous PPP on the $\mathbb{R}^2$ plane with a density of $\lambdaris$. Each RIS is equipped with $L$ reflecting elements (REs), aiding in the redirection and boosting of signals. It is assumed that HAPs and users use isotropic antennas for simplicity.

The user U is assumed to receive signals from the nearest HAP through the nearest RIS and the direct HAP–U link.
\subsection{Signal Model}
The received signal at the ground user can be expressed as 
\begin{align}
\label{eq:Y}
y &= A\,s+w= \bigg(\frac{\sum_{l=1}^{L} \,q_{l}\, g_{l}\,\exp(j\theta_{l})}{R_q^{\frac{\epsilon_q}{2}}\,\rg^{\frac{\epsilon_g}{2}}}+ \frac{u}{\ru^{\frac{\epsilon_u}{2}}}\bigg)\,s+w,
\end{align}
where $s$ is the transmitted signal, $q_l$ and $g_l$ are the fading coefficients for HAP--RIS and RIS--U links, respectively, with $R_q$ and $R_g$ representing the corresponding distances, and $\epsilon_q$ and $\epsilon_g$ representing the corresponding path-loss exponents. The additive white Gaussian noise, $w$, has zero mean and variance of $N_0 = \E[|w|^2]$. Additionally, $\theta_l$ represents the adjustable phase induced by the $l$th RE. Assuming coherent signal combining and beamforming toward the intended user, interference from other HAPs via RISs is negligible. However, the user may still experience interference directly from other HAPs, which can be investigated in future work.

The maximum instantaneous signal-to-noise ratio (SNR) at U is obtained by setting $\theta_{l}=\angle{u}-\left(\angle{q_{l}}+\angle{g_{l}}\right)$ as
\begin{align}
\label{eq:snr}
  \rho&=\rho_0\,|A|^2=\rho_0\,\left(\frac{\sum_{l=1}^{L} |q_{l}\, g_{l}|}{R_q^{\frac{\epsilon_g}{2}}\,\rg^{\frac{\epsilon_q}{2}}}+\frac{|u|}{\ru^{\frac{\epsilon_u}{2}}}\right)^2, 
\end{align}
where $\rho_0=E_s/N_0$ is the transmit SNR with $E_s=\E[|s|^2]$. 

\subsection{Distance Distributions}
Given the stochastic nature of the HAP and RIS locations, the horizontal distances $\omega_g$, $\omega_q$, and $\omega_u$ of the RIS--U, HAP--RIS, and HAP--U links are random variables. A performance assessment requires deriving the distribution of these horizontal distances, in addition to key statistical measures of the corresponding diagonal distances $\rg$, $R_q$, and $\ru$.

In particular, the probability density function (PDF) of the horizontal distance $\omega$, where $\omega \in \{\omega_g, \omega_q, \omega_u\}$ denotes the distances from an arbitrarily located user $U$ to the nearest RIS, from the nearest RIS to the nearest HAP, or from $U$ to the nearest HAP, is given by~\cite{ppp_distribution}
\begin{align}\label{eq:pdf}
f_{\omega}(w) &= 2\lambda\pi w \exp{(-\lambda\pi w^2)},
\end{align}
for $w \geq 0$. Here, the shorthand $\lambda \in \{\lambda_{\text{RIS}}, \lambda_{\text{HAP}}\}$ represents both RISs and HAPs densities jointly.

Based on (\ref{eq:pdf}), the $t^{\mathrm{th}}$ moment of the related measure $R^{\frac{-\epsilon}{2}}$ is calculated as
\begin{align}\label{eq:mean_rqq}
\E\Big[R^{\frac{-t\,\epsilon}{2}}\Big]&=2\lambda\pi  \int_{0}^{\infty}{ (w^2+H^2)^{-\frac{t\,\epsilon}{4}}}\nonumber\\
& \:\:\:\:\:\:\times w\exp{(-\lambda\pi w^2)}\, \mathrm{d} w\nonumber\\
&=\Big(\pi\,\lambda\Big) ^{\frac{t\,\epsilon }{4}} e^{\pi  H^2 \lambda}\,  \Gamma \left(1-\frac{t\,\epsilon }{4},H^2 \lambda\pi \right),
\end{align}
where $\Gamma(\cdot, \cdot)$ denotes the upper incomplete Gamma function~\cite{tableofseries}. The parameters $\epsilon \in \{\epsilon_g, \epsilon_q, \epsilon_u\}$ and $R=\sqrt{\omega^2+H^2} \in \{R_g, R_q, R_u\}$ represent the path-loss exponents and the diagonal distances for the RIS--U, HAP--RIS, and HAP--U links, respectively. The height $H$ is calculated as
\begin{equation}
H =
\begin{cases}
\HHAP - \HRIS & \text{for the HAP--RIS link}, \\
\HRIS & \text{for the RIS--U link}, \\
\HHAP & \text{for the HAP--U link}.
\end{cases}
\end{equation}

\begin{IEEEproof}
Expression (\ref{eq:mean_rqq}) follows from the definition of the $t^{\mathrm{th}}$ moment of a random variable. The integral is evaluated using \cite[3.383.8]{tableofseries}.
\end{IEEEproof}

\begin{figure*}[t]
\setcounter{MYtempeqncnt}{\value{equation}}
\setcounter{equation}{11}
\begin{equation}
\resizebox{.85\hsize}{!}{$\begin{aligned}
\label{eq:mean_A}
&
\left.\begin{aligned} \E[|A|]= &\overbrace{\frac{L\,\pi}{4}\sigma_g\,\sigma_q\,{\mathrm{e}^{-K_q}}\left(\frac{ m_g}{m_g+K_g}\right)^{m_g} { }_1F_1\left(1+\frac{1}{2} ; 1 ; K_q\right){ }_2 F_1\left(1+\frac{1}{2}, m_g, 1, \frac{K_g}{ m_g+K_g}\right) }^{\E[\xi]}\Big(\pi\,\lambdaris\Big)^{\frac{\epsilon_g }{4}}\\
 &\hspace{-0.03 cm}\times \Big(\pi\,\lambdahaps\Big) ^{\frac{\epsilon_q }{4}} \, e^{\pi  \HRIS^2 \lambdaris}\,  
  e^{\pi  (\HHAP - \HRIS)^2 \lambdahaps}\, \Gamma \left(1-\frac{\epsilon_g }{4},\HRIS^2 \lambdaris\pi \right) \Gamma \left(1-\frac{\epsilon_q }{4},(\HHAP - \HRIS)^2 \lambdahaps\pi \right)
  \end{aligned}\right\}\text{$P_1$} \\
 &\hspace{1.3cm} +\underbrace{\sigma_u\Gamma\left(1+\frac{1}{2}\right)\Big(\pi\,\lambdahaps\Big) ^{\frac{\epsilon_u }{4}} e^{\pi  \HHAP^2 \lambdahaps}\,  \Gamma \left(1-\frac{\epsilon_u }{4},\HHAP^2 \lambdahaps\pi \right)}_{P_2}
\end{aligned}$}
\end{equation}
\setcounter{equation}{\value{MYtempeqncnt}}
\hrulefill
\end{figure*}
\begin{figure*}[t]
\setcounter{MYtempeqncnt}{\value{equation}}
\setcounter{equation}{14}
\begin{equation}
\resizebox{0.95\hsize}{!}{$\begin{aligned}
\label{eq:var_A}
\Var[|A|]= &\left[\left(L\,\sigma_g^2\sigma_q^2+(L^2-L)\frac{\pi^2}{16}\,{\mathrm{e}^{-2K_q}}\left(\frac{ m_g}{m_g+K_g}\right)^{2m_g} { }_1F_1^2\left(1+\frac{1}{2} ; 1 ; K_q\right){ }_2 F_1^2\left(1+\frac{1}{2}, m_g, 1, \frac{K_g}{ m_g+K_g}\right) \right) \Big(\pi\,\lambdaris\Big)^{\frac{\epsilon_g }{2}}\Big(\pi\,\lambdahaps\Big) ^{\frac{\epsilon_q }{2}}\right.\\
 & \times \left.e^{\pi  \HRIS^2 \lambdaris} \,
  e^{\pi  (\HHAP - \HRIS)^2 \lambdahaps}\, \Gamma \left(1-\frac{\epsilon_g }{2},\HRIS^2 \lambdaris\pi \right) \Gamma \left(1-\frac{\epsilon_q }{2},(\HHAP - \HRIS)^2 \lambdahaps\pi \right) -{P_1^2}\Bigg)\right]\\
& +\Bigg[ {\sigma_u^2\Big(\pi\,\lambdahaps\Big) ^{\frac{\epsilon_u }{2}} e^{\pi  \HHAP^2 \lambdahaps}\,  \Gamma \left(1-\frac{\epsilon_u }{2},\HHAP^2 \lambdahaps\pi \right)}-P_2^2\Bigg]. 
\end{aligned}$}
\end{equation}
\setcounter{equation}{\value{MYtempeqncnt}}
\hrulefill\\
\begin{small}
$^*$Note: $P_1$ and $P_2$ in (\ref{eq:var_A}) are defined in (\ref{eq:mean_A}).
\end{small}
\end{figure*}
\subsection{Fading Model}
Given that both the serving HAP and the serving RIS are positioned at elevated locations, a LoS path is expected for both RIS hops. Therefore, Rician fading is the most suitable model for the HAP--RIS and RIS--U channels. In addition, shadowing attenuation due to signal blockage by obstacles surrounding the user is considered in our analysis to affect the RIS--U link only. To model this, we assume composite shadowed-Rician fading for the RIS--U link, capturing both LoS and shadowing effects and Rician fading for the HAP--RIS link. 
In particular, the $t$th moment of a composite shadowed-Rician fading coefficient, $g_l$, of the RIS--U link is given by~\cite{simplified-shadowed-rician}  
\begin{align} 
\label{eq:shadowed-mean}
E\left[|g_l|^t\right]= & \left(\frac{2 b_g m_g}{2 b_g m_g+\Omega_g}\right)^{m_g}\left(2 b_g\right)^{\frac{t}{2}} \Gamma\left(1+\frac{t}{2}\right)\nonumber \\ 
& \times{ }_2 F_1\left(1+\frac{t}{2}, m_g, 1, \frac{\Omega_g}{2 b_g m_g+\Omega_g}\right)
\end{align}
where $\Omega_g$ is the average power of the LoS component, $2b_g$ is the average power of the multi-path component excluding the LoS, and $m_g$ is the Nakagami parameter of the RIS--U link.

We can rewrite (\ref{eq:shadowed-mean}) in terms of the Rician factor $K_g$, which is the ratio between the power in the LoS component and the power in the other scattered paths, i.e., $K_g=\frac{\Omega_g}{2b_g}$
as 
\begin{align} 
\label{eq:shadowed-mean_K}
E\left[|g_l|^t\right]= &\frac{\left(\sigma_g^2\right)^{\frac{t}{2}}}{(K_g+1)^{\frac{t}{2}}} \left(\frac{m_g}{m_g+K_g}\right)^{m_g} \Gamma\left(1+\frac{t}{2}\right)\nonumber \\ 
& \times{ }_2 F_1\left(1+\frac{t}{2}, m_g, 1, \frac{K_g}{m_g+K_g}\right)
\end{align}
where $\sigma_{g}^2 =\E\big[|g_l|^2\big]=2b_g(1+K_g)$ is the average gain of the envelope of the fading coefficient $g_l$.

The $t$th moment of a Rician distributed fading coefficient, $q_l$, of the HAP--RIS link is derived from (\ref{eq:shadowed-mean_K}) by taking the limit $\lim_{m_g\to\infty}E\left[|g_l|^t\right]$. This results in the following expression 
\begin{align}
\label{eq:rician-mean}
\mathbb{E}\left[|q_l|^t\right]=\frac{\left(\sigma_q^2\right)^{\frac{t}{2}}\Gamma\left(1+\frac{t}{2}\right) \mathrm{e}^{-K_q}}{\left(K_q+1\right)^{\frac{t}{2}}} { }_1F_1\left(1+\frac{t}{2} , 1 , K_q \right).  
\end{align}
The Rician factor $K_q=\frac{\Omega_q}{2b_q}$ represents the ratio of the power in the LoS component $\Omega_q$ to the power in the other scattered paths $2b_q$ for the HAP--RIS link, whereas $\sigma_{q}^2 =\E\big[|q_l|^2\big]$. The above formula can also be acquired from \cite[Eq. 3]{kappa_mu_2} by setting $\kappa=K_q$ and $\mu=1$. 


Furthermore, the direct HAP--U path is assumed to follow Rayleigh distribution, which models the scenario where no LoS exists between the serving HAP and the user. The $t$th moment of a Rayleigh-distributed fading coefficient, $u$, can be found by substituting $m_g=0$ in (\ref{eq:shadowed-mean_K}) to yield \cite{Islam_RISs}
\begin{align}
\label{eq:rayleigh-mean}
 \E[|u|^t]=(\sigma_u^2)^{\frac{t}{2}}\, \Gamma\left(1+\frac{t}{2}\right),
\end{align}
where $\sigma_{u}^2 =\E\big[|u|^2\big]$ represents the average power of the scattered multi-path components for the direct link.

\section{Performance Analysis}
This section derives the coverage probability and ergodic capacity of the RIS-based HAPs network. In particular, the system's performance evaluation requires first the derivation of the PDF of the end-to-end SNR.

\subsection{Statistical Channel Characterization Based on Gamma
Distribution}

The combined channel response $|A|$ in (\ref{eq:snr}) is approximately normally distributed, and its PDF resembles a Gaussian distribution with a single peak and truncated tail on the left side, as shown in Fig.~\ref{fig:pdf}. This Gaussian distribution can be further approximated by a Gamma distribution, as described in \cite{stochastic-book}, with the PDF given by
\begin{align}
\label{eq:pdf_A}
    f_{|A|}(x) &\simeq \frac{x^{\alpha-1}}{\beta^{\alpha}\,\Gamma(\alpha)} \exp\left(-\frac{x}{\beta}\right).
\end{align}
Using this approximation, the PDF of the end-to-end SNR, defined in (\ref{eq:snr}) can be computed with the relation \(f_{\rho}(x) = \frac{1}{2 \sqrt{\rho_0 x}} f_{|A|}\left(\sqrt{\frac{x}{\rho_0}}\right)\), resulting in
\begin{align}
\label{eq:pdf_snr}
    f_{\rho}(x) &\simeq \frac{1}{2\,\beta^{\alpha}\,\Gamma(\alpha)} \rho_0^{-\frac{\alpha}{2}} x^{\frac{\alpha-2}{2}} \exp\left(-\sqrt{\frac{x}{\beta^2\, \rho_0}}\right),
\end{align}
where \(\alpha = \frac{(\E[|A|])^2}{\Var[|A|]}\) and \(\beta = \frac{\Var[|A|]}{\E[|A|]}\). 

The mean \(\E[|A|]\) in (\ref{eq:pdf_snr}) is calculated using its linearity property together with the independency assumption as $\E[{|A|}]=\E[\xi]{\E[R_q^{-\frac{\epsilon_q}{2}}]\E[\rg^{\frac{-\epsilon_g}{2}}]}+{\E[|u|]}{\E{[\ru^{-\frac{\epsilon_u}{2}}]}}$, where $\E[\xi]=\E\big[\sum_{l=1}^{L} |q_{l}\, g_{l}|\big]=L\,\E\big[|q_{l}|]\E[|g_{l}|\big]$. This yields  $\E[|A|]$ as given in (\ref{eq:mean_A}) at the beginning of the previous page.\stepcounter{equation}
Similarly, the variance can be calculated as
\begin{align}
 &\Var[|A|]=\Var\left[\sum_{l=1}^{L} |q_{l}\, g_{l}| R_q^{-\frac{\epsilon_q}{2}}\rg^{\frac{-\epsilon_g}{2}}\right]+\Var\left[\ru^{-\frac{\epsilon_u}{2}}|u|\right]\nonumber\\
 &=\E\big[\sum_{l=1}^{L} \big|q_{l}\, g_{l}\big|^2\big]\E[R_q^{-{\epsilon_q}}]\E[R_g^{-{\epsilon_q}}]-L^2\E[|q_{l}|\big]^2\E[|g_{l}|\big]^2\nonumber\\
 &\times\E[R_q^{-\frac{\epsilon_q}{2}}]^2\E[R_g^{-\frac{\epsilon_q}{2}}]^2+(\E[\ru^{-{\epsilon_u}}]\E[|u|^2]
-\E[\ru^{-\frac{\epsilon_u}{2}}]^2\E[|u|]^2)
\end{align}
with 
\begin{align}
&\E\left[\sum_{l=1}^{L}| q_{l}\, g_{l}|^2\right]=\Var\left[\sum_{l=1}^{L} |q_{l}\, g_{l}|\right]+\E\left[\sum_{l=1}^{L} |q_{l}\, g_{l}|\right]^2\nonumber\\
&=L\,\Var\big[\big|q_{l}\, g_{l}\big|\big]+L^2\,\E[|q_{l}|\big]^2\E[|g_{l}|\big]^2\nonumber\\
&=L\,\E\big[\big|q_{l}\big|^2\big]\E\big[\big| g_{l}\big|^2\big]+(L^2-L)\,\E\big[|q_{l}|\big]^2\E\big[|g_{l}|\big]^2,
\end{align}
which leads to evaluating $\Var[|A|]$ as shown in (\ref{eq:var_A}) at the top of the previous page.

\stepcounter{equation}
\subsection{Coverage Probability} 
If the user connects to the nearest HAP and is assisted by the nearest RIS, the coverage probability can be calculated as\cite[Eq.~11]{tanash-RIS_sat}
\begin{align}
\label{eq:cov_prob}
\operatorname{P}_c&\simeq1-\frac{\gamma\left(\alpha,\sqrt{\frac{\rho_\mathrm{th}}{\rho_0\beta^2}}\right)}{\Gamma(\alpha)},
\end{align}
where $\alpha$ and $\beta$ are calculated using the mean and variance of the combined channel response as given in (\ref{eq:mean_A}) and (\ref{eq:var_A}), respectively.

\subsection{Ergodic Capacity}
The ergodic capacity of the considered system shares the same analytical form as given in \cite[Eq.~12]{tanash-RIS_sat}. It is restated herein as
\begin{equation}
\resizebox{1.0\hsize}{!}{$\begin{aligned}
\label{eq:capacity}
&\bar{\C}\simeq\frac{\pi}{\alpha\Gamma(\alpha)\log_e(2)}\bigg(\frac{1}{\beta^2 \rho_0}\bigg)^{{\frac{\alpha}{2}}}\csc\bigg(\frac{\pi\alpha}{2}\bigg)\, _1F_2\left(\frac{\alpha}{2};\frac{1}{2},1+\frac{\alpha}{2};-\frac{1}{4\beta^2\rho_0}\right)\\
&+\frac{\Gamma(\alpha-2)}{\Gamma(\alpha)}\frac{1}{\beta^{2}\rho_0}\, _2F_3\left(1,1;2,\frac{3}{2}-\frac{\alpha}{2},2-\frac{\alpha}{2};-\frac{1}{4\beta^2\rho_0}\right)-\frac{1}{(1 + \alpha)}\\
&\times \frac{1}{\Gamma(\alpha)}\Bigg[(2-\alpha-2\alpha^2+\alpha^3)\Gamma(\alpha-2)\Big(\log\Big(\frac{1}{\beta^2\rho_0}\Big)-2\psi^{(0)}(\alpha)\Big)\\
&+\pi\bigg(\frac{1}{\beta^2\rho_0}\bigg)^{\frac{1+\alpha}{2}}\,_1F_2\left(\frac{1}{2}+\frac{\alpha}{2};\frac{3}{2},\frac{3}{2}+\frac{\alpha}{2};-\frac{1}{4\beta^2\rho_0}\right)\sec\Big(\frac{\pi\alpha}{2}\Big)\Bigg].
 \end{aligned}$}
\end{equation}
where $\psi^{(0)}(\cdot)$ denotes the 0-th polygamma function and $\csc(\cdot)$ represents the cosecant function \cite{TCOM2}. This formulation incorporates the novel parameters $\alpha$ and $\beta$ derived in this work.

\begin{figure}
    \centering
    \includegraphics[width=.485\textwidth]{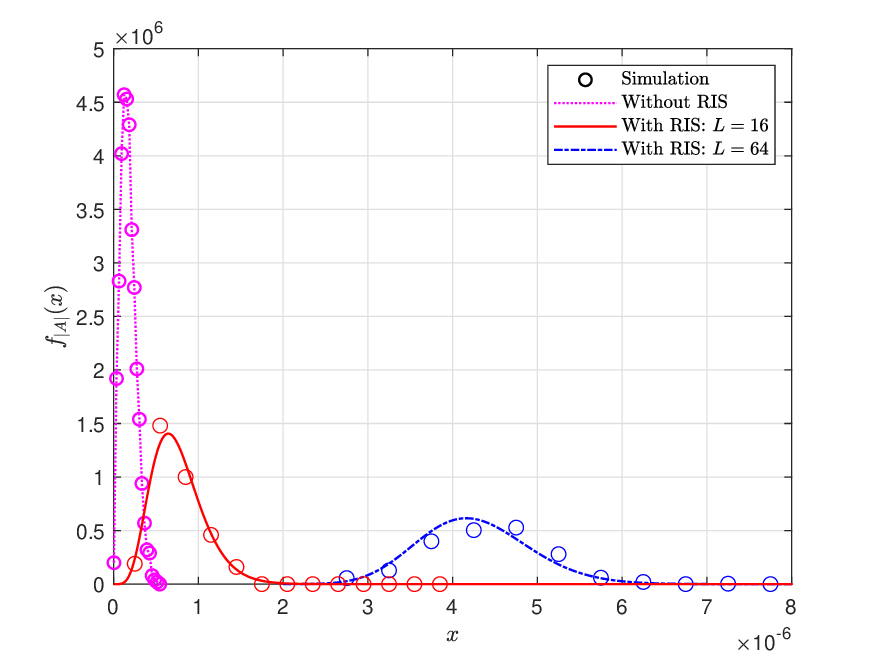}
    \caption{The PDF of the combined channel response $|A|$ for $L=16$ and $L=64$.}
    \label{fig:pdf}
\end{figure}

%
\section{Results and Observations}
This section presents a numerical validation of the adopted Gamma-distributed SNR approximation using Monte Carlo simulations. We also evaluate the system's performance by analyzing the derived coverage probability and ergodic capacity and comparing them with the corresponding reference simulations. Thus, in the figures, analytical results are represented by lines, while circular markers denote simulated results.

In particular, we consider herein a network setup with $\lambda_{\text{HAP}} = 5 \times 10^{-7} / \mathrm{m}^{2}$, $\lambda_{\text{RIS}} = 5 \times 10^{-4} / \mathrm{m}^{2}$, $H_{\mathrm{HAP}} = 50~\mathrm{km}$, $H_{\mathrm{RIS}} = 50~\mathrm{m}$, $\epsilon_g = 2$, $\epsilon_q = 3$, $\epsilon_u = 3$, and for the link budget $E_s = 10 \,\mathrm{W}$, and $N_0=-92 \,\mathrm{dBm}$. We consider two scenarios that reflect distinct environmental conditions for the normalized shadowed-Rician fading in the RIS–U link, where $\sigma_{g}^2 = 1$, and the normalized Rician fading in the HAP--RIS link, where $\sigma_{q}^2 = 1$.
\begin{itemize}
    \item Scenario 1: Frequent heavy shadowing (FHS) environment with $K_g = 0.0071$ and $m_g = 0.739$ for the RIS–U link, combined with severe fading (SF) for the HAP–RIS link, characterized by $K = 0.1$ \cite{simplified-shadowed-rician}. This configuration models the worst-case scenario.

   \item Scenario 2: Infrequent light shadowing (ILS) environment with $K_g = 4.0823$ and $m_g = 19.4$ for the RIS–U link, while the HAP–RIS link maintains weak fading (WF) environment with $K = 10$, representing an enhanced-fading scenario.
\end{itemize}
The HAP--U link is assumed to follow normalized Rayleigh fading with $\sigma_{u}^2 =1$ in both scenarios.

The accuracy of the Gamma approximation in (\ref{eq:pdf_A}) for the combined channel response $|A|$ in (\ref{eq:snr}) of the considered system model is tested and illustrated in Fig.~\ref{fig:pdf}. It can be noted that the developed approximation is very tight, and the high accuracy is maintained for low and high numbers of REs in the RIS. For comparison, the communication scenario with only the HAP--U link is also illustrated, revealing that the integration of RIS significantly enhances power gain. Furthermore, this gain increases with higher values of $L$, as evidenced by the rightward shift of the PDF.

\begin{figure}
    \centering
    \includegraphics[width=.485\textwidth]{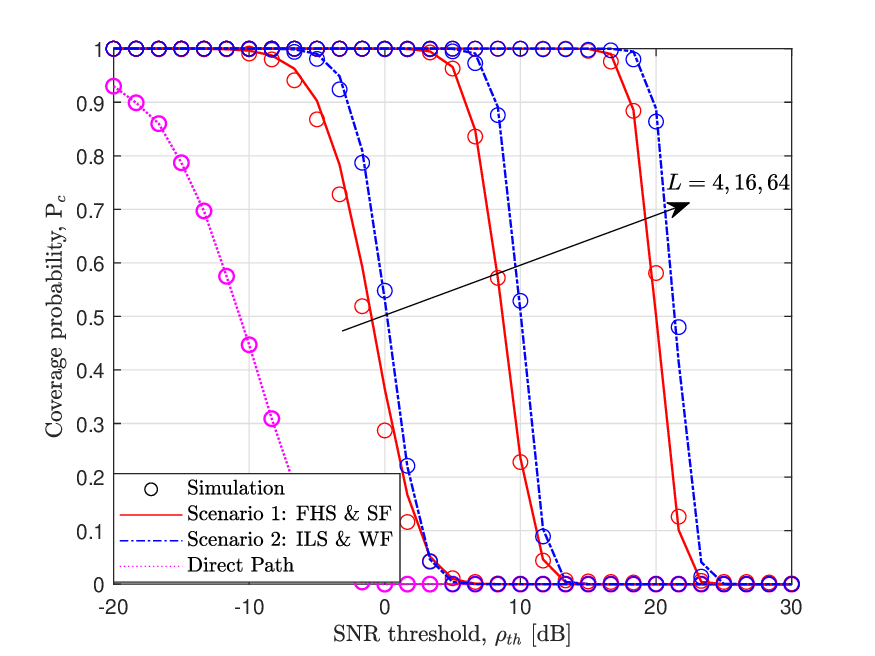}
    \caption{Coverage probability comparison for direct HAP--U link versus RIS-assisted communication with HAPs under different fading and shadowing scenarios.}
    \label{fig:p_c}
\end{figure}

The coverage probability of the considered system is depicted in Fig. \ref{fig:p_c}, further validating the high accuracy of the Gamma-approximated PDF across different fading and shadowing environments where the analytical results closely match the simulated measurements. Coverage probabilities for both fading scenarios are compared, demonstrating that integrating RIS into the system significantly enhances coverage performance. This improvement is observed across both environmental conditions compared to the scenario where communication relies solely on the direct path between the HAP and the user. The RIS enhances signal reliability in challenging environments and amplifies performance in more favorable conditions. Additionally, increasing the number of REs further boosts the system's performance, highlighting the effectiveness of RIS in various propagation scenarios. For example, to achieve a coverage probability of $50\%$, the required SNR threshold increases from $0$ dB to $10$ dB for both scenarios when the number of REs rises from $L=4$ to $L=16$, indicating that increasing the number of REs enhances the signal strength significantly, enabling the system to maintain the same coverage probability at a higher SNR threshold. 

\begin{figure*}[ht]
\begin{center}
\subfigure[]
{\includegraphics[trim=0cm 0cm 0cm 0cm, clip=true, width=0.49\textwidth ]{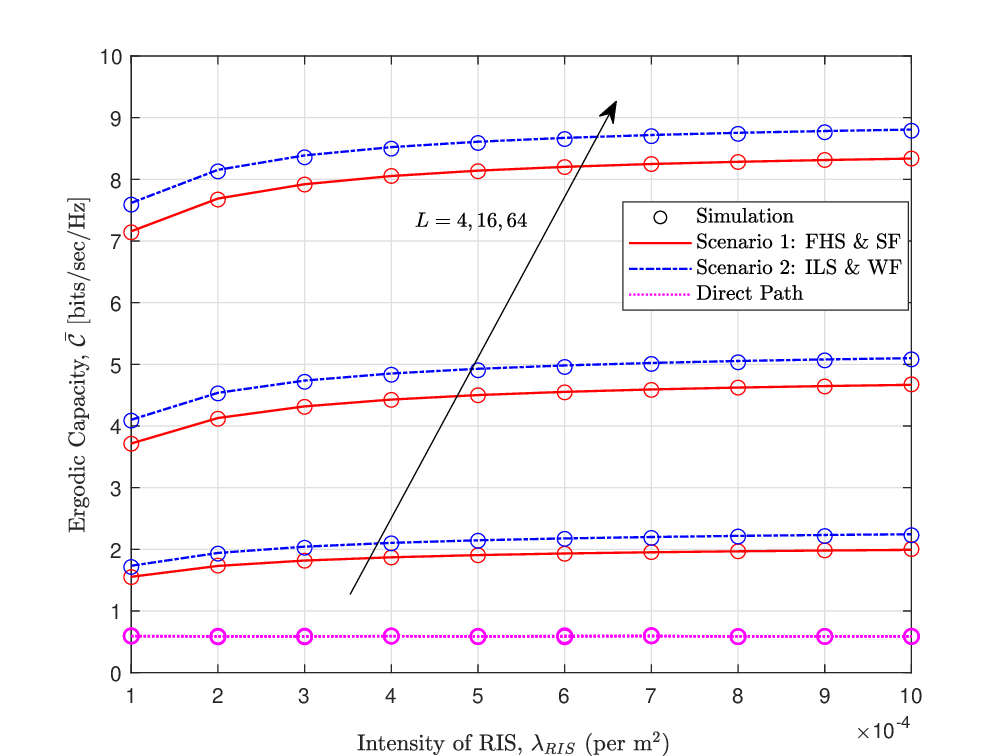}}
\subfigure[]{\includegraphics[trim=0cm 0cm 0cm 0cm, clip=true, width=0.49\textwidth ]{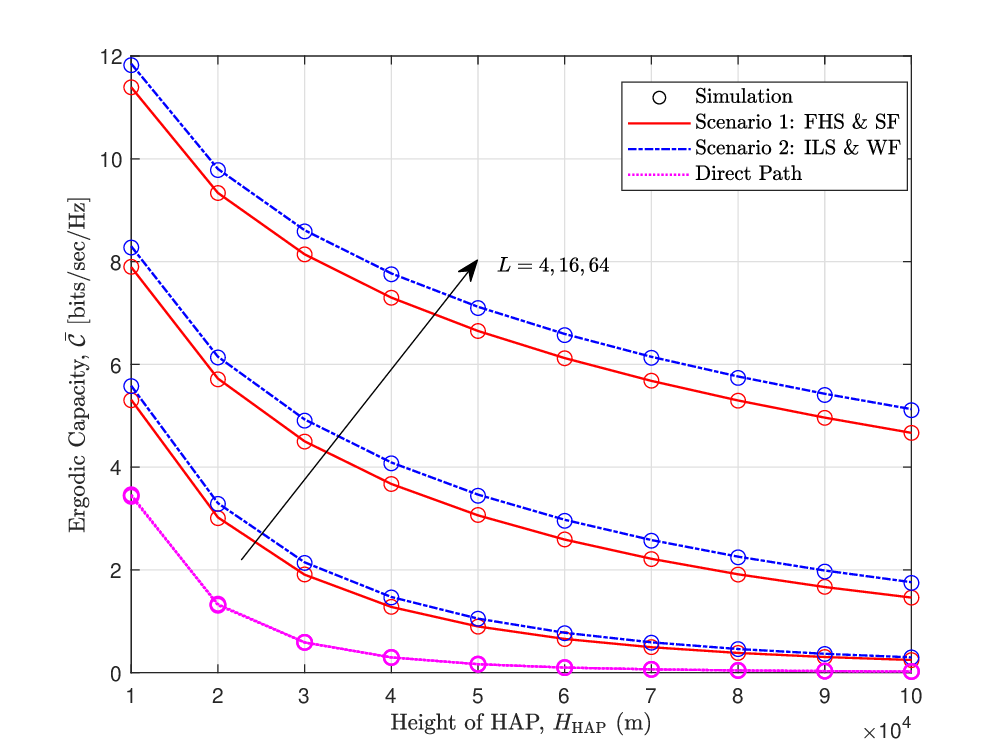}}
 \caption{\textit{Impact of RIS configuration on performance:} (a) The ergodic capacity vs.\ the intensity of the RIS, $\lambda_{\text{RIS}}$; (b) The ergodic capacity vs.\ the height of the HAP, $\HHAP$, for different numbers of REs, 
 $L$.}
\label{fig:ergodic_capacity}
\end{center}
\end{figure*}

Fig. \ref{fig:ergodic_capacity} illustrates the impact of the intensity of RIS deployment and the height of the HAP on capacity performance for different numbers of REs. As the density of RIS deployment increases, the distance to the nearest RIS decreases, leading to improved performance and increased ergodic capacity as depicted in Fig. \ref{fig:ergodic_capacity}(a). However, once a certain RIS density is reached, the system saturates for each $L$ value such that additional RIS units no longer enhance capacity. This occurs because the environment is sufficiently populated with RIS units to maintain optimal coverage.

As expected, the performance of the RIS-assisted system decreases with higher HAP altitudes, as shown in Fig. \ref{fig:ergodic_capacity}(b), where lower capacity is achieved. This decline occurs because increasing the HAP height results in longer signal propagation distances, causing more significant path loss and reducing the effectiveness of the RIS in enhancing signal strength. 

\section{Conclusion}
This paper demonstrated the advantages of incorporating RISs into HAP networks. Stochastic geometry was used to model the spatial distribution of HAPs and RISs. Through analytical derivations and Monte Carlo simulations, we showed that RIS effectively reduces the effects of shadowing and path loss, improving coverage probability and ergodic capacity. The performance gains are more noticeable with higher RIS element counts and denser deployments, although saturation occurs beyond a certain threshold. The study also shows that increasing HAP altitude negatively impacts system performance due to higher path loss. These results offer a robust foundation for designing and optimizing RIS-enhanced HAP systems in diverse deployment scenarios. Future work will focus on interference modeling and resource management in large-scale deployments.

\bibliographystyle{IEEEtran}
\bibliography{IEEEabrv,refs}

\end{document}